\documentclass[epj]{webofc}
\usepackage[utf8]{inputenc}
\usepackage[varg]{txfonts}   
\usepackage{booktabs}
\usepackage{xcolor}
\usepackage{subfigure} 
\definecolor{darkred}{rgb}{0.4,0.0,0.0}
\definecolor{darkgreen}{rgb}{0.0,0.4,0.0}
\definecolor{darkblue}{rgb}{0.0,0.0,0.4}
\usepackage[bookmarks,linktocpage,colorlinks,
    linkcolor = darkred,
    urlcolor  = darkblue,
    citecolor = darkgreen]{hyperref}
\usepackage{subfigure}
\wocname{EPJ Web of Conferences}
\woctitle{Lattice2017}
\newcommand{\code}[1]{\mbox{\texttt{\detokenize{#1}}}}

\DeclareFixedFont{\ttb}{T1}{txtt}{bx}{n}{9} 
\DeclareFixedFont{\ttm}{T1}{txtt}{m}{n}{9}  

\usepackage{color}
\definecolor{deepblue}{rgb}{0,0,0.5}
\definecolor{deepred}{rgb}{0.6,0,0}
\definecolor{deepgreen}{rgb}{0,0.5,0}

\usepackage{listings}

\begin{document}
%
\selectlanguage{english}
\title{%
	Automated lattice data generation
}
\author{%
	\firstname{Venkitesh} \lastname{Ayyar}\inst{1} \and
	\firstname{Daniel~C.}  \lastname{Hackett}\inst{1}\fnsep%
	\thanks{Speaker, \email{daniel.hackett@colorado.edu}} \and
	\firstname{William~I.}  \lastname{Jay}\inst{1} \and
	\firstname{Ethan~T.}  \lastname{Neil}\inst{1,2}
}
\institute{%
	Department of Physics, University of Colorado, Boulder, Colorado 80309, USA
	\and
	RIKEN-BNL Research Center, Brookhaven National Laboratory, Upton, New York 11973, USA
	\and
	Raymond and Beverly Sackler School of Physics and Astronomy, Tel~Aviv University, 69978 Tel~Aviv, Israel
}
\abstract{%
	The process of generating ensembles of gauge configurations (and measuring various observables over them) can be tedious and error-prone when done ``by hand''. In practice, most of this procedure can be automated with the use of a workflow manager. We discuss how this automation can be accomplished using Taxi, a minimal Python-based workflow manager built for generating lattice data. We present a case study demonstrating this technology.
}
\maketitle
\section{\texttt{taxi}: Production workflow management for lattice gauge theory}
\label{sec:intro}

Workflow management in any large-scale calculation using high-performance computing may present a significant logistical challenge.  Efficient use of computational resources is usually best accomplished by dividing the overall project into a number of smaller tasks, which can then be carried out in parallel by a pool of worker jobs.  Since the tasks can have complicated inter-dependencies, coordination of the workers is essential when distributing tasks from the available pool.

For lattice gauge theory (LGT) in particular, most projects share a common task dependency structure which arises from the Markov-Chain Monte Carlo (MCMC) methods central to such calculations.  First, an ensemble of configurations is generated in a Markov chain.  Each configuration-generation task depends on the previous configuration-generation task in the chain.  Second, physical observables are measured over the ensemble.  To make a measurement on any given configuration, that configuration must already exist, so each measurement task depends on a configuration-generation task.  This gives the overall workflow a tree-like structure: the ``trunk'' gauge ensemble must be completed serially, but once it is finished multiple other tasks can branch off from each configuration-generating task.

Given this structure, automating the workflow to generate a single LGT ensemble can be relatively straightforward; one can use a simple self-resubmitting script that will evolve the gauge configurations one-by-one, and then similar scripts which will run through the set of available configurations serially and compute any observables of interest.  However, this approach is inflexible: with every worker specialized to a single task, each worker must exit when work for its specialization is exhausted, regardless of the state of the rest of the task pool.  On high-performance computers with significant competition in the job queue, this can increase the wall time required to complete a set of tasks.  Moreover, managing each ensemble with individual scripts can become cumbersome when the number of ensembles and/or observables becomes large, a common situation in modern LGT calculations.

These considerations motivate the use of a \emph{workflow manager}, which can flexibly assign new tasks to general-purpose worker jobs as old tasks are completed.  This allows a worker which exhausts a certain type of task, or which reaches the end of a branch of tasks, to pick up and continue running different types of tasks, or to start working on another branch.  For projects with many ensembles and complicated sets of tasks, controlling and modifying a single centralized task repository is much simpler than interacting with a disparate set of scripts.  The existence of a central ``task plan'' with information about dependencies and status also enables the automation of the overall workflow, including verification of task completion and error recovery.

Our collaboration became interested in automation and workflow management to enable our exploration of the thermodynamics of SU(4) gauge theory with fermions in multiple representations \cite{chiral_proceedings,confinement_proceedings}.  
In the course of our investigation, we ran $\sim 4\times10^5$ HMC trajectories over $10^3$ ensembles, measuring both Wilson flow and spectroscopy for both representations of fermion.
Generating this volume of data without significant automation proved logistically intractable, leading us to explore workflow managers.

There are a number of general workflow management systems available already, such as Pegasus~\cite{pegasus}, Makeflow~\cite{makeflow}, Apache~Taverna~\cite{taverna}, and Kepler~\cite{kepler}.  We chose to create a new tool, \code{taxi}, for two main reasons, both related to ease of use for our particular problem.
First, \code{taxi} is specialized to the MCMC-focused workflow of lattice gauge theory; this is a less general approach, but as a trade-off requires less code to be written for different MCMC calculations.  Our second motivation was the requirement, typical in large-scale lattice calculations,of running the same project on a number of different remote machines, with heterogenous software environments, network access restrictions, etc.  Setting up one of the more general workflow managers listed above (and all of their dependencies) on many different systems where user access is limited can present a significant (or even impassable) logistical obstacle. \code{taxi} is designed to be lightweight and flexible, requiring only Python 2.6.6 (included with most Linux distributions) as a minimal dependency and modularized to work with different queueing systems interchangeably.

In the next section, we give an overview of \code{taxi}'s design philosophy for task management in general and for structure MCMC workflow plans in particular.  Section~\ref{sec:example} gives an example of basic usage for a toy LGT workflow.  Finally, in section~\ref{sec:status} we give an overview of the current status and near-future plans for developing \code{taxi}.

\section{System architecture}
\label{sec:arch}

\subsection{Task management model}

The \code{taxi} workflow system has two types of actor: Taxis and the Dispatcher.  A Taxi is a general-purpose worker which executes computing tasks assigned by the Dispatcher.  The Dispatcher contains the planned workflow in full, including priorities, dependencies, and completion status for all tasks.  Whenever a Taxi completes a task, it queries the Dispatcher to receive the next available task of highest priority; the Dispatcher provides the appropriate task and updates its records.

A key feature of this ``Taxi-Dispatcher'' model is that the Dispatcher is only active when it is queried by a Taxi.  This allows the Dispatcher in \code{taxi} to be structured as a passive repository, typically built on top of some form of SQL database, with Dispatcher-specific logic built in to the interface to the database seen by the Taxis.  This removes the need for any additional monitor program to be run on the remote machine, instead using the workers to carry out organizational tasks.  Contrast this with the more typical ``overlord-minion'' model, in which central coordination is provided by an active monitor program which tracks the workers and assigns new tasks when it detects that workers are idle.  In the latter model, the ``overlord'' program must be active or accessible on the remote machine at all times, which can be difficult to maintain reliably (dropped connections, etc.), potentially wasteful of computational resources (if run on compute nodes), and/or irritating to administrators (if a resource-intensive monitor is run on an access or compile node).

In practice, we divide the passive central repository in to the Dispatcher, which contains information about tasks only, and the Pool, which tracks the status of the Taxis in the queue as well as their resource limits (maximum time and number of nodes).  This separation allows for easier adjustment of the desired number of workers and resource limits, independent of the set of available tasks, and for easier coordination when running multiple projects simultaneously on the same queue.

\subsection{MCMC workflow structure}
\label{sec:interface}

As discussed in section~\ref{sec:intro}, there is a common structure to MCMC workflows, regardless of the exact theory under investigation or software suite being used.
\code{taxi} takes advantage of this structure to minimize the work necessary to specify data generation tasks (i.e., to specify a ``run'') and to adapt the software to a new application.
This comprises most of the user-facing component of \code{taxi}, as most or all of workflow planning and task management for typical workflows is taken care of transparently and automatically, without any additional input from the user.

\code{taxi} includes two abstract superclasses for running MCMC tasks.
Instances of \code{ConfigGenerator} run binaries that advance the Markov chain and generate new configurations: in LGT this is usually a Hybrid Monte Carlo (HMC) binary.
Instances of \code{ConfigMeasurement} run binaries that perform measurements on existing configurations: some LGT examples include measuring correlation functions (spectroscopy) or computing various observables as the configuration is evolved under Wilson flow.
Adapting \code{taxi} to a new application or lattice software suite (e.g., a variant of MILC) amounts to implementing subclasses of these two abstract classes.
This process requires the user to implement only a few methods in each subclass: a constructor that preprocesses and stores the parameters needed to run the binary; a \code{build_input} method that generates the input string or input file to be fed to the binary, using the parameters stored in each object; and a \code{verify_output} method that checks that the output of the binary is present, well-formed, and complete.
To give a sense of scale, the five subclasses in our MILC application are each $100-300$ lines of low-density Python code.

Most of the rest of the necessary functionality to generate MCMC data is built in to the abstract superclasses.
For example: each sequence of gauge configurations in a run will start with either a fresh start (i.e., a random configuration or a unit configuration), from a configuration generated by a previous run and stored in the filesystem, or by branching off from another sequence of configurations.
The \code{ConfigGenerator} superconstructor takes a single \code{starter} parameter that detects and handles each of these cases appropriately.
Similarly, the \code{ConfigMeasurement} constructor takes a \code{measure_on} parameter that may be used to specify either a stored gauge configuration file or a \code{ConfigGenerator} task instance from the same run.
If \code{starter} or \code{measure_on} is used to specify a \code{ConfigGenerator} from the same run, the object will read relevant physical parameters from the specified \code{ConfigGenerator} instance, obviating the need to specify them by hand (and removing the opportunity to specify mismatched parameters).
If \code{starter} or \code{measure_on} is instead used to specify a stored file, \code{taxi} uses modularized file naming conventions (easily specified by the user and automatically detected by the software) to read physical parameters from the filename, similarly simplifying the process of making measurements (any necessary parameters not present in the filename must still be specified by the user).
This behavior is helpful in specifying long chains of \code{ConfigGenerator} tasks and even more useful in the case of measurements: for example, to measure correlation functions on some configuration, one need only specify the parameters specific to spectroscopy (e.g., smearing radius and boundary conditions) versus having to carefully provide all the matching parameters used to generate the configuration.

\section{Usage example}
\label{sec:example}

Figure~\ref{fig:run-spec} shows the code for a working toy example which sets up and launches two jobs (i.e., two worker taxis) which coordinate to generate two ensembles of pure gauge data on $4^4$ lattices and apply the Wilson flow to those ensembles.  This example illustrates a number of convenience functions which further simplify run specification for the user.

In the first block of code in the \code{__main__} section of the script, the \code{make_config_generator_stream} convenience function is used to construct two chains of instances of \code{PureGaugeORATask}, a subclass of \code{ConfigGenerator} that runs a pure-gauge MCMC binary.
The first set, \code{seed_stream}, will generate an ensemble of 10 configurations at $\beta=7.75$ from a fresh start, with each stored configuration separated by 100 trajectories.
The second set, \code{fork_stream}, will generate a second ensemble of 5 configurations at $\beta=7.76$. To cut down on equilibration time versus a fresh start, \code{fork_stream} forks off from \code{seed_stream} after the fifth configuration.
Integrator parameters (such as the number of overrelaxation steps or number of quantum heatbath steps to run per trajectory) are specified by default arguments to the PureGaugeORATask constructor.

In the second block, all fifteen \code{PureGaugeORATask}s are pooled in to \code{cg_pool}, and the \code{measure_on_config_generators} convenience function is used to specify the measurement of Wilson Flow across both ensembles.
This convenience function creates instances of \code{config_measurement_class} applied to all \code{ConfigGenerator} instances supplied to the argument \code{measure_on}.
Each instance of \code{config_measurement_class}, in this case \code{FlowTask}, steals parameters from the \code{ConfigGenerator} instances they are associated with and so only the flow integrator step size \code{epsilon} and the maximum time to flow \code{tmax} need to be provided to perform the measurement.
The parameter \code{start_at_traj} specifies that the first two configurations (i.e. 200 trajectories) are for equilibration and thus not to be measured on; consequently, \code{flow_pool} contains only eleven instances of \code{FlowTask}.

The remainder of the data generation process is set in to motion by the final two blocks.
In the third block, the run-specification script initializes a \code{Dispatcher}, which compiles the task pool in to a SQLite dispatch database to be stored in the provided location.
The fourth block of code generates two taxis, each of which are to run on one node; instantiates a \code{Pool} object for the taxis to coordinate among themselves; and registers those taxis with both the \code{Dispatcher} and \code{Pool}.
The final line launches the appropriate number of taxis to start the job working: in this case, one taxi will be launched, as there is one ``trunk'' job available to work on at the beginning (the first task in \code{seed_stream}).
Once launched, the taxis will automatically maintain the optimal number of running taxis (equal to the number of active sequences of \code{ConfigGenerator}s, or ``trunk number'' of the task pool): so, after the fifth task in \code{seed_stream} completes, a second taxi will be launched to work on \code{fork_stream} in parallel.
After launching the first taxi, the run-specification script exits and (barring task failures) no further user intervention is required.

\begin{figure}
\begin{lstlisting}[
	language=Python,
	basicstyle=\ttm,
	otherkeywords={self, with, if},
	keywordstyle=\ttb\color{deepgreen},
	emph={},          % Custom highlighting
	emphstyle=\ttb\color{deepblue},    % Custom highlighting style
	stringstyle=\color{deepred},
	frame=lines,
	]
import taxi
import taxi.mcmc
from taxi.pool import SQLitePool
from taxi.dispatcher import SQLiteDispatcher
from taxi.apps.milc.pure_gauge_ora import PureGaugeORATask
from taxi.apps.milc.flow import FlowTask

PureGaugeORATask.binary = './pg_ora'
FlowTask.binary = './flow'

if __name__ == '__main__':
    seed_stream = taxi.mcmc.make_config_generator_stream(
        config_generator_class=PureGaugeORATask,
        starter=None,
        req_time=240, streamseed=1, N=10, 
        Ns=4, Nt=4, beta=7.75, n_traj=100
    )
    fork_stream = taxi.mcmc.make_config_generator_stream(
        config_generator_class=PureGaugeORATask,
        starter=seed_stream[4],
        req_time=240, streamseed=2, N=5,
        Ns=4, Nt=4, beta=7.76, n_traj=100
    )

    cg_pool = seed_stream + fork_stream
    flow_pool = taxi.mcmc.measure_on_config_generators(
        config_measurement_class=FlowTask,
        measure_on=cg_pool,
        req_time=60, start_at_traj=200,
        tmax=1, epsilon=.03
    )

    job_pool = cg_pool + flow_pool
    my_disp = SQLiteDispatcher(db_path="./dispatch.sqlite")
    my_disp.initialize_new_job_pool(job_pool)
	
    taxi_list = [taxi.Taxi(time_limit=10*60, nodes=1)) for i in range(2)]	
    my_pool = SQLitePool(db_path="./pool.sqlite",
        work_dir="./work/", log_dir="./log/")
    for my_taxi in taxi_list:
        my_pool.register_taxi(my_taxi)
        my_disp.register_taxi(my_taxi, my_pool)

    my_pool.spawn_idle_taxis(dispatcher=my_disp)
\end{lstlisting}
\caption{Example run-specification script. This script specifies and launches a run that will generate two ensembles of pure-gauge data, including a measurement of the Wilson flow.}
\label{fig:run-spec}
\end{figure}

\section{Current status and future plans}
\label{sec:status}

An experimental release of the taxi software package is available on GitHub \cite{taxi}.
The GitHub repository includes an example application (a suite of runners for our custom multi-representation variant of the MILC binary suite, which should be easily adaptable to any MILC derivative), and a number of examples/templates for common use cases, such as generating new ensembles of configurations while performing standard measurements, or performing measurements on an existing set of stored configuration files.
These examples should be sufficient to illustrate basic use.
In the near future, we hope to provide detailed documentation that elaborates on how to adapt \code{taxi} to new applications, how to use \code{taxi} for common use cases, and how to adapt \code{taxi} to run on new clusters or machines.
Also in development is a suite of toy applications and examples, so that users won't need to first adapt \code{taxi} to their binary suite to play with example workflows.

The driving design requirement of \code{taxi} has been ease-of-use.
For ease of installation, the software is designed to work transparently with Python's \code{virtualenv} virtual environment system, which allows users to install Python software painlessly on remote machines without administrator rights.
In the context of getting \code{taxi} up and running, \code{virtualenv} provides an application-sufficient subset of the capabilities of container systems like Docker with the benefit of being nearly trivial to set up and install without enhanced user privileges or help from administrators.
Localizations (i.e., the framework required to run on different supercomputers or different queueing systems) in \code{taxi} are modular, meaning that \code{taxi} can be adapted to run on new machines with a minimum of coding.
The current distribution of taxi includes localizations for an SGE-based queueing system and for the USQCD machines at Fermilab, which use TORQUE PBS.
Coming soon are localizations for machines that use SLURM, as well as for the Redis queueing system, a local queue to run on workstations and laptops.
Another possible queueing backend is METAQ \cite{Berkowitz:2017vcp, Berkowitz:2017xna}, which would allow \code{taxi} to manage tasks inside a larger bundled job.

The package includes a set of command-line tools to monitor job progress and to recover from common failure cases.
For example, it is often the case that HMC jobs are initially run with overly-optimistic integrator parameters or estimated walltime requirements, resulting in task failure or job cancellation.
The package includes tools to adjust task parameters in situ and to roll back failed tasks.
For more sophisticated monitoring and modification of active runs, \code{taxi} and its tool suite have been designed with an eye towards using the Jupyter notebook as a dashboard interface.
Examples of this use case are forthcoming.

Currently, Dispatcher and Pool are implemented in SQLite, a local file-based implementation of SQL.
While SQLite is adequate for running on single machines, a central networked database will be necessary to coordinate runs across multiple machines (as well as providing other benefits).  To this end, we are looking in to adapting \code{taxi} to other flavors of SQL, particularly the open-source \mbox{PostgreSQL}.
Because much of the Dispatcher logic is implemented abstractly and modularly, adaptation to other types of database would be equally straightforward.

To push forward the automation horizon, we have been considering what is possible when automated analysis of automatically-generated data is used to specify further data to generate, thus ``closing the loop'' on data generation.
This idea leads straightforwardly to the complete automation of several tedious tasks traditionally performed by humans.
Among these tasks are scaling tests on supercomputers, tuning of HMC integrator parameters, and running data until some threshold amount of statistics has been achieved.
An even more sophisticated and physically interesting case is in thermodynamics: it should be possible to completely automate the exploration of bare-parameter phase diagrams.
In such a case, the user need only specify the ranges of parameters to explore (including a bounding set of conditionals, e.g. ``the AWI quark mass is less than a certain value'').
In this closed pipeline, \code{taxi} generates data; automated analysis software measures observables on the data as it is produced (e.g., the thermodynamic phase, how long it took an ensemble to equilibrate, the AWI quark mass); finally, a further piece of automated run-specification software examines the analyzed data and adds new tasks to the ongoing run to continue the exploration.
Such a closed pipeline can be used to precisely (to whatever specified threshold) pin down the location of finite-temperature transitions or $\kappa_c$ lines with only minimum human intervention.
The only piece of this loop that we have not yet implemented in practice is the automated run-specification component.

\subsection*{Acknowledgements}

\footnotesize
Our research was supported in part by the U.S.~Department of Energy under grant number DE-SC0010005.  Brookhaven National Laboratory is supported by the U.~S.~Department of Energy under contract DE-SC0012704.

\bibliography{lattice2017-taxi}

\end{document}